\documentclass[reqno,a4paper,oneside,11pt]{amsart}

\usepackage[usenames,dvipsnames]{xcolor}
\usepackage[colorlinks=true,linkcolor=Blue,citecolor=Green]{hyperref}
\usepackage{amsmath,amssymb,epsf}
\usepackage{amsfonts,amsbsy,amscd,stmaryrd}
  \SetSymbolFont{stmry}{bold}{U}{stmry}{m}{n}
\usepackage{latexsym,amsopn}
\usepackage{amsthm}
\usepackage{esint}
\usepackage{mathrsfs}
\usepackage{slashed} 
\usepackage{enumitem}
\renewcommand{\baselinestretch}{1.05}
\usepackage[margin=1.3in,footskip=0.25in]{geometry}
\allowdisplaybreaks

\usepackage[utf8]{inputenc}
\usepackage[T1]{fontenc}   

\usepackage{graphicx}
\usepackage{multirow}
\usepackage{float}

\newcounter{smallarabics}
\newenvironment{arabicenumerate}
{\begin{list}{{\normalfont\textrm{(\arabic{smallarabics})}}}
  {\usecounter{smallarabics}\setlength{\itemindent}{0cm}
   \setlength{\leftmargin}{5ex}\setlength{\labelwidth}{4ex}
   \setlength{\topsep}{0.75\parsep}\setlength{\partopsep}{0ex}
   \setlength{\itemsep}{0ex}}}
{\end{list}}

\newcounter{smallroman}

\newcommand{\ben}{\begin{arabicenumerate}}  
\newcommand{\een}{\end{arabicenumerate}}

\newtheorem{theorem}{Theorem}[section]

\newtheorem{proposition}[theorem]{Proposition}
\newtheorem{lemma}[theorem]{Lemma}

\theoremstyle{definition}
\newtheorem{definition}[theorem]{Definition}
\newtheorem{remark}[theorem]{Remark}
\newtheorem{example}[theorem]{Example}

\newcommand{\beq}{\begin{equation}}
\newcommand{\eeq}{\end{equation}}
\newcommand{\bea}{\begin{aligned}}
\newcommand{\eea}{\end{aligned}}
\newcommand{\bear}{\begin{array}{rl}}
\newcommand{\eear}{\end{array}}
\newcommand{\bex}{\begin{example}}
\newcommand{\eex}{\end{example}}
\def\bel{\begin{lemma}}
\def\eel{\end{lemma}}
\def\bet{\begin{theoreme}}
\def\eet{\end{theoreme}}
\def\bed{\begin{definition}}
\def\eed{\end{definition}}
\def\ber{\begin{remark}}
\def\eer{\end{remark}}
\def\bep{\begin{proposition}}
\def\eep{\end{proposition}}
\newcommand{\qeds}{\qed\medskip}

\let\origmaketitle\maketitle
\def\maketitle{
  \begingroup
  \def\uppercasenonmath##1{} 
  \let\MakeUppercase\relax 
	\origmaketitle
  \endgroup
}

\def\rr{{\mathbb R}}
\def\zz{{\mathbb Z}}
\def\cc{{\mathbb C}}

\def\ss{{\mathbb S}}

\def\bar{\overline}

\def\cinf{C^\infty}
\def\proof{
\noindent{\bf Proof.}\ \ }

\usepackage{calrsfs}
\DeclareMathAlphabet{\pazocal}{OMS}{zplm}{m}{n}

\def\cL{{\pazocal L}}
\def\cS{{\pazocal S}}

\def\cN{{\pazocal N}}

\def\cX{{\pazocal X}}

\def\sH{\mathcal{H}}

\def\sI{\mathcal{I}}

\def\CAR{{\rm CAR}}
\def\wf{{\rm WF}}

\def\i{{\rm i}}

\def\sgn{{\rm sgn}}
\def\id{{\rm id}}

\DeclareMathOperator{\Ker}{Ker}

\DeclareMathOperator{\supp}{supp}
\DeclareMathOperator{\sing}{sing}
\DeclareMathOperator{\Char}{Char}

\def\p{\partial}

\def\14{\frac{1}{4}}

\def\12{\frac{1}{2}}

\def\e{{\rm e}}

\newcommand{\one}{\boldsymbol{1}}

\def\sc{{\rm sc}}
\def\sgn{{\rm sgn}}
\def\id{{\rm id}}

\def\coinf{C_{\rm c}^\infty}

\DeclareSymbolFont{boldoperators}{OT1}{cmr}{bx}{n}
\SetSymbolFont{boldoperators}{bold}{OT1}{cmr}{bx}{n}

\def\t{{\scriptscriptstyle\#}}

\makeatletter
\newcommand*{\defeq}{\mathrel{\rlap{%
                     \raisebox{0.34ex}{$\m@th\cdot$}}%
                     \raisebox{-0.4ex}{$\m@th\cdot$}}%
                     =}
                     
\makeatother
\makeatletter
\newcommand*{\eqdef}{=\mathrel{\rlap{%
                     \raisebox{0.34ex}{$\m@th\cdot$}}%
                     \raisebox{-0.4ex}{$\m@th\cdot$}}%
                     }
\makeatother

\def\Sol{{\rm Sol}}

\def\WF{{\rm WF}}

\def\Vol{{\rm vol}}

\def\dual{\!\cdot \!}
\def\kst{^{*}}
\def\stk{{}^{*}\!}

\def\DD{{\mathbb D}}

\def\SS{{\mathbb S}}
\def\CAR{{\rm CAR}}

\def\mo{\mathit{o}}

\def\zero{{\mskip-4mu{\rm\textit{o}}}}

\def\dVol{\mathop{}\!d{\rm vol}}
\def\diff{\mathop{}\!d}
\def\End{\mathit{End}}

\def\bS{\mathbb{S}}

\def\MI{{\rm M}_{\rm I}}
\def\MII{{\rm M}_{\rm II}}
\def\MIp{{\rm M}_{\rm I'}}
\def\MK{{\rm M}}
\def\MIUII{{\rm M}_{{\rm I}\cup {\rm II}}}
\def\2Sol{{\rm Sol}_{{\rm L}^{2}}}

\newcommand{\open}[1]{\mathopen{}\mathclose{\left]#1 \right[}}

\newcommand{\opencl}[1]{\mathopen{}\mathclose{\left]#1 \right]}}

\def\MIII{{\rm M}_{\rm III}}
\newcommand{\abs}[1]{{\left\vert #1 \right\vert}}

\numberwithin{equation}{section}

\begin{document}

\title[Hadamard property of the Unruh state in the large a case]{\Large Hadamard property of 
the Unruh state for massless fermions on Kerr spacetime : the large $a$ case}

\author{Dietrich \textsc{H\"afner}}
\address{Universit\'e Grenoble Alpes, Institut Fourier, 100 rue des Maths, 38610 Gi\`eres, France}
\email{dietrich.hafner@univ-grenoble-alpes.fr}

\author{Christiane \textsc{Klein}}
\address{Universit\'e Grenoble Alpes, Institut Fourier, 100 rue des Maths, 38610 Gi\`eres, France\\ CY Cergy Paris Universit\'e, 2 avenue Adolphe Chauvin, 95302 Cergy-Pontoise, France}
\email{christiane.klein@univ-grenoble-alpes.fr}
\keywords{Quantum Field Theory on curved spacetimes, Hadamard states, Dirac equation, Kerr spacetime, Hawking temperature}
\subjclass[2010]{81T13, 81T20, 35S05, 35S35}
\thanks{\emph{Acknowledgments.} We thank Stefan Hollands and Micha\l\,  Wrochna for helpful comments on an earlier version of this paper. We also acknowledge support from the ANR funding  ANR-20-CE40-0018-01.}

\begin{abstract} 
In \cite{GHW}, the Unruh state for massless fermions on a Kerr spacetime was constructed and the authors showed its Hadmard property in the case of very slowly rotating black holes $\abs{a}\ll M$. In this note, we extend this result to the full non extreme case $\abs{a}<M$. 
\end{abstract}

\maketitle
\section{Introduction}
When studying quantum fields in black hole spacetimes, one crucial and non-trivial step is the choice of state for the quantum field at hand. Ideally, the state should be motivated by the physical situation. One state which satisfies this condition for a variety of black hole spacetimes is the Unruh state \cite{Un} - it appears as the final state when one considers the collapse of a star to a black hole, see e.g. \cite{Haw}, \cite{Ha}.

Moreover, the selected state should be physically reasonable, in the sense of satisfying the Hadamard property, see e.g. \cite{FV} for why this is the necessary condition. In this note, we will consider the microlocal formulation of the Hadamard property due to Radzikowski \cite{Ra}. The Unruh state is also favourable in this regard. It was first constructed and its Hadamard property shown for the wave equation on the Schwarzschild spacetime by Dappiaggi et at. \cite{DMP}, and this proof has since been expanded to other spherically symmetric black hole spacetimes \cite{BJ,HWZ}.

However, astrophysical black holes are expected to be rotating, and therefore not spherically symmetric. Rotating black holes are described by the Kerr spacetime. Constructing explicit Hadamard states in the case of the Kerr spacetime is more complicated because of the lack of a global timelike Killing vector field in the exterior region. Nonetheless, G\'erard, Wrochna and the first author have recently constructed the Unruh state for massless Dirac fields on the Kerr spacetime in \cite{GHW} (see \cite{Ha} for an earlier construction of the Unruh state on block I also in the massive charged case). They have also shown its Hadamard property, but only under the condition that the angular momentum per unit mass $a$ of the black hole is small. 
In \cite{Ha}, \cite{GHW}, to circumvent the lack of a global timelike Killing field, the data are separated in data for solutions going to the black hole horizon and solutions going to infinity by means of asymptotic velocities constructed in the framework of scattering theory, see \cite{HN}. Two different Killing vector fields are then used to construct the state. Later on, the second author constructed the Unruh state for the Klein-Gordon equation on the De Sitter Kerr spacetime in \cite{Kl} and showed its Hadamard property, again under the condition that the angular momentum per unit mass of the black hole is small. The scattering construction is there replaced by an argument using the fast decay of the field (see e.g. \cite{Dy2}, \cite{HV}).

The assumption of small angular momentum is an important shortcoming, since most black holes are expected to rotate rapidly, see \cite{An} and \cite{Th}. The restriction is used in \cite{GHW} to show the Hadamard property. For sufficiently small $a$, any geodesic in the black hole exterior $\MI$ (see Section \ref{secKerr} for the definition) either reaches the past black hole horizon or past null infinity or meets a region, where both vector fields which have been used to construct the Unruh state are timelike. For those geodesics who reach either the past black hole horizon or past null infinity, a refinement of a strategy of Moretti \cite{Mo}, see also \cite{Ho}, \cite{GW}, showing a Hadamard property from characteristic data was used. This argument works the same way in the large $a$ case. Yet, for large $a$ not all other geodesics meet a region where both relevant vector fields are timelike.

In this note, we extend the result of \cite{GHW} to all subextreme values of the angular momentum $\abs{a}<M$.  This requires a precise analysis of the situation on the backward/forward trapped set. The argument therefore only concerns the geodesic flow and should in principle apply also to other fields. Note, however, that in the case of the De Sitter Kerr metric other problems arise, in particular in the full range of angular momentum, it is up to today unknown if the wave equation has growing modes or not.
It should be pointed out that the Unruh state in block I is constructed for massive charged Dirac fields on the Kerr-Newman metric in \cite{Ha}. This uses scattering results of \cite{Da}. As long as the field has no mass and charge we expect that the results of this paper, including the Hadamard property, continue to hold also on the Kerr-Newman metric. In the massive charged case, however, the geometric interpretation in terms of solutions of a characteristic Cauchy problem at infinity is expected to fail, and a new analysis would be required. Concerning the extreme case, let us mention the scattering results of Borthwick for the extreme De Sitter Kerr case \cite{Bo}. Similar results are expected to hold for the extreme Kerr case, but no geometric interpretation of the results is given in \cite{Bo}.
We refer to \cite{GHW} for a full review of the literature.    

The paper is organized as follows. In Section \ref{secDirac}, \ref{secKerr}, and \ref{secTraces}, we revise the necessary constructions to define the Unruh state in Section \ref{secmain}, where the main result of this paper is stated. Sections  \ref{secDirac}-\ref{secTraces} essentially summarizes the constructions in \cite{GHW}, we also refer to \cite{GHW} for more details and additional references. In Section \ref{secproof}, we prove our main result.

\section{The Dirac and Weyl operators}
\label{secDirac}
In this section, we summarize some elementary facts about Dirac and Weyl operators. For details of the construction, we refer to \cite{GHW}.  

\subsection{Notation} 
Following the notation of \cite{GHW}, if $(M, g)$ is an oriented and time oriented Lorentzian manifold of dimension $4$, we denote by $\Omega_{g}\in \Lambda^{4}M$ the volume form associated to $g$ and by $\dVol_{g}= | \Omega_{g}|$ the volume density.

If $S$ is a smooth hypersurface of $M$, we denote by $i: S\to M$ the canonical injection.

Given an oriented, time-oriented Lorentzian manifold $(M, g)$, we denote by $(M', g)$ the same Lorentzian manifold with the opposite time orientation. That is, $\id : (M, g)\to (M', g)$ is an isometric involution reversing the time orientation. Given that $(M, g)$ admits a spinor bundle $\cS\xrightarrow{\pi}M$, so does $(M', g)$. Objects associated to $(M', g)$ will generally be decorated by a prime. 

The space of smooth sections of a vector bundle $\cS\xrightarrow{\pi}M$ is denoted by $\cinf(M;\cS)$. If $\cS$ is a complex vector bundle, its dual bundle is denoted by $\cS^\#$, its anti-dual bundle by $\cS^*$, and the complex conjugate bundle (obtained by considering the fibres as complex vector spaces with the opposite complex structure) by $\bar{\cS}$.

If $\cX$ is a complex vector space, and $\beta\in  L(\cX,\cX^*)$ a sesquilinear form, we write $\overline{\psi_1}\cdot \beta \psi_2$ for its evaluation on $\psi_1,\psi_2\in  \cX$. If instead  $\beta\in \cinf(M;L(\cS,\cS^*))$, the  fibre-wise evaluation on  $\psi_1,\psi_2\in  \cinf(M;\cS)$ is written in the same way.

\subsection{Dirac and Weyl operators} 
Let $(M,g)$ be a globally hyperbolic spacetime of dimension 4, equipped with its unique spin structure. From the spin structure, one obtains in a canonical way a spinor bundle $\cS\xrightarrow{\pi}M$ of rank $4$, a spin connection $\nabla^{\cS}$, and a representation $\gamma$ of the Clifford bundle ${\rm Cl}(M, g)$ in $End(\cS)$. One also obtains a positive energy Hermitian form $\beta$, and a complex conjugation $\kappa$ acting on the fibres of $\cS$.  
The  massless Dirac operator 
$\slashed{D}$  acting on smooth sections of the canonical spinor bundle $\cS$ over $M$ is the differential operator defined as
\[
\slashed{D}= g^{\mu\nu}\gamma(e_{\mu})\nabla^{\cS}_{e_{\nu}},
\]
where $(e_{0}, \dots ,e_{3})$ is a local frame of $TM$. In the massless case, it is well known that the whole analysis can be reduced to the \emph{Weyl equation} 
\begin{equation}
\label{e10.9}
\DD \phi=0,
\end{equation}
which accounts for half of the degrees of freedom. We call $\DD$ the associated Weyl operator. 
More precisely, one can identify $\cS$ with $\SS^{*}\oplus \SS^{\t}$, where $\SS^{*}=\overline{\SS^{\t}}$ is the bundle of even Weyl spinors.  We recall that 
\beq\label{e10.6c}
\Gamma(X)= \beta \gamma(X)\in\cinf(M, \End(\SS^{*}, \SS)), \ \ X\in \cinf(M; TM)
\eeq
and 
\beq\label{e10.8}
\DD= g^{\mu\nu}\Gamma(e_{\mu})\nabla_{e_{\nu}}^{\cS}.
\eeq
 If $S$ is a space-like Cauchy surface, the Cauchy problem
\begin{equation}
\label{e4.0}
\begin{cases}
\DD \phi= 0, \\
r_{S}\phi= \varphi\in \coinf(S; \SS^{*}_{S}),
\end{cases}
\end{equation}  where $\SS^{*}_{S}$ is the restriction of $\SS^{*}$ to $S$ and  $r_{S}\phi= \phi_{| S}$,  has a unique solution $\phi\eqdef \mathbb{U}_{S}\varphi$ in the space of smooth space-compact solutions, $\Sol_{\rm sc}(M)$. $\phi$ \emph{space-compact} means that the intersection of $\supp\phi$ with any space-like Cauchy surface is compact.

The \emph{principal symbol} of  $\DD$ is the section $\sigma_\DD\in\cinf(T^*M\setminus \zero;L(\SS^*,\SS))$ given by
\beq\label{eq:sigmadd}
\sigma_\DD(x,\xi)=\Gamma(g^{-1}(x)\xi), \ \ (x,\xi)\in T^*M\setminus \zero.
\eeq
By \cite[Lemma 3.1]{GHW}, the Weyl operator $\DD$ is pre-normally hyperbolic, meaning one can find a differential operator ${\mathbb{D}'}$ such that $(\sigma_{\DD}\circ\sigma_{{\DD}'})(x,\xi)=(\xi\cdot g^{-1}(x)\xi) \one$. 
As a consequence, the \emph{characteristic manifold} of $\DD$ defined as
\[
\Char(\DD)=\{ (x,\xi)\in T^*M\setminus\zero \,:\, \sigma_\DD(x,\xi) \mbox{ is not invertible} \}.
\]
is given by  
\[
\Char(\DD)=\{ (x,\xi)\in T^*M\setminus\zero \,:\,  \xi\cdot g^{-1}(x)\xi = 0 \}\eqdef \cN,
\]
the null cone in $T^*M$. It has two connected components, the future and past null cones
\begin{equation}
\label{sloubi1}
\cN^{\pm}\defeq\cN\cap \{(x, \xi)\in T^*M\setminus\zero \,:\, \pm v\dual \xi>0\ \forall v\in T_{x}M\hbox{ future directed timelike}\}.
\end{equation}
If $v \dual \xi>0$ for all $v\in T_{x}M$ which are future directed timelike, we say that $\xi$ is future pointing. 

\subsection{$L^2$ solutions}
As in \cite{GHW}, we will denote the space of smooth space-compact solutions of $\DD\phi=0$, $\phi\in \cinf(M; \SS^{*})$, by $\Sol_{\rm sc}(M)$. 
Then, for any $\phi_1$, $\phi_2\in \Sol_{\rm sc}(M)$, the  {\em current}  $J(\phi_{1}, \phi_{2})\in \cinf(M; T^{*}M)$ defined by
\beq\label{e10b.0}
J(\phi_{1}, \phi_{2})\dual X\defeq \bar{\phi}_{1}\dual \Gamma(X)\phi_{2},\ \ X\in \cinf(M; TM),
\eeq
is conserved:
\begin{equation}
\label{e10b.00}
\nabla^{a}J_{a}(\phi_{1}, \phi_{2})=0.
\end{equation}
The current conservation, the identity $\nabla^{a}J_{a}\Omega_{g}= d (g^{-1}J\lrcorner \Omega_{g})$, and an application of Stokes' formula yield
\beq\label{stoker}
\int_{\p {U}}i^{*}(g^{-1}J(\phi_{1}, \phi_{2})\lrcorner \Omega_{g})=0, \ \phi_{i}\in \Sol_{\rm sc}(M),
\eeq
if $U$ is any open set whose boundary $\p U$ is a union of smooth hypersurfaces, such that $\supp J(\phi_{1}, \phi_{2})\cap \p U$ is compact. 

Another useful way to write an expression of the form 
\[
\int_{S}i^{*}(g^{-1}J\lrcorner \Omega_{g})
\]
for $J$ a $1$-form on $M$ and $S\subset M$ a smooth hypersurface, is by choosing a future pointing vector field $l= l^{a}$ transverse to $S$, and a $1$-form $\nu= \nu_{a}dx^{a}$ on $M$ such that $TS= \Ker \nu$, and $\nu\dual l=1$ to obtain
\beq\label{corr.e-1}
i^{*}(g^{-1}J\lrcorner\Omega_{g})= (\nu\dual g^{-1}J)i^{*}(l\lrcorner \Omega_{g}).
\eeq

We set 
\beq\label{e4.02}
\bea
(\phi_{1}| \phi_{2})_{\DD}&\defeq \i\int_{S} i^{*}(g^{-1}J(\phi_{1}, \phi_{2})\lrcorner \Omega_{g})\\[2mm]
&=\i\int_{S}\bar{\phi}_{1}\dual \Gamma(g^{-1}\nu)\phi_{2} \mathop{}\,i^{*}_{l}d{\rm vol}_{g},
\eea
\eeq
where $i^{*}_{l}d{\rm vol}_{g}= |i^{*}(l\lrcorner \Omega_{g})|$, and $S\subset M$ is now any smooth, not necessarily space-like, Cauchy surface. It follows from \eqref{stoker} that the r.h.s.~in \eqref{e4.02} is in fact independent of the choice of the Cauchy surface $S$. 

If $S$ is space-like with future pointing normal $n$, this can be written as
\beq\label{trif.1}
(\phi_{1}| \phi_{2})_{\DD}= \i \int_{S} \bar{\phi_{1}}\dual \Gamma(n)\phi_{2} \dVol_{h},
\eeq
where $\dVol_h=i^{*}_{n}d{\rm vol}_{g}$. By the properties of $\beta$ described in \cite{GHW}, $\i \Gamma(n)$ is positive definite, which shows that 
 $(\cdot| \cdot)_{\DD}$ is  a Hilbertian scalar product on $\Sol_{\rm sc}(M).$ We can reformulate this as being a scalar product on the Cauchy data on a spacelike hypersurface $\Sigma$ defined by
\begin{align}
\label{scalarproduct}
\overline{\varphi}_{1}\dual\nu_{\Sigma}\varphi_{2}=\i \int_{\Sigma} \bar{\varphi}_{1}\dual \Gamma(n)\varphi_{2}\,d\Vol_{h}, \ \varphi_{i}\in \coinf(\Sigma; \SS^{*}_{\Sigma}).
\end{align}

\begin{definition}{\cite[Definition 3.2]{GHW}}\label{def4.1}
The Hilbert space   $\Sol_{{\rm L}^{2}}(M)$, called the {\em space of $L^{2}$ solutions}, is  the completion of $\Sol_{\rm sc}(M)$ for the scalar product $(\cdot| \cdot)_{\DD}$.
 \end{definition}
As noted in \cite{GHW}, there is a continuous embedding $\Sol_{{\rm L}^{2}}(M)\subset L^{2}_{\rm loc}(M; \SS)$ implying that elements of $\Sol_{{\rm L}^{2}}(M)$ are distributional solutions of $\DD \phi=0$.
\subsection{Action of Killing vector fields}
Let $X$ be a complete Killing vector field on $(M, g)$, and denote by $\cL_X$ the corresponding Lie derivative on spinors. Then $\cL_{X}$ preserves $\Sol_{\rm sc}(M)$. 
One has 
\begin{proposition}{\cite[Proposition 3.5]{GHW}}\label{prop10.1}
The operator $\i^{-1}\cL_{X}$ with domain  $\Sol_{\rm sc}(M)$ is essentially self-adjoint on the Hilbert space $\Sol_{{\rm L}^{2}}(M)$. 
\end{proposition}

\section{The Kerr spacetime}
\label{secKerr}
Here, we summarize the relevant facts on the Kerr black hole geometry collected in \cite[Section 5]{GHW}.

\subsection{Boyer-Lindquist blocks} 
Following \cite{GHW}, we set
\[
\begin{array}{l}
\Delta = r^{2}- 2Mr+a^2,\ \ 
\rho^{2} = r^{2}+ a^{2}\cos^{2}\theta,\\[2mm]
 \sigma^{2} = (r^{2}+a^{2})^{2}- a^{2}\Delta\sin^{2}\theta= (r^{2}+a^{2})\rho^{2}+ 2 a^{2}Mr\sin^{2}\theta.
\end{array}
\]
To restrict ourselves to the \emph{subexetreme} Kerr case, we fix an $0<\abs{a}<M$. As a consequence, $\Delta$ as a function of $r$ has two distinct, positive roots $r_{\pm}= M\pm \sqrt{M^{2}- a^{2}}$.

In this setting, the three Boyer-Lindquist blocks are the manifolds $(\MI, g)$, $(\MII, g)$, and $(\MIII,g)$ given by
\[
\begin{array}{l}
\MI = \rr_{t}\times\open{r_{+},+\infty}_r\times\mathbb{S}^{2}_{\theta, \varphi},\\[2mm]
\MII = \rr_{t}\times\open{r_{-},r_{+}}_r\times\mathbb{S}^{2}_{\theta, \varphi},\\[2mm]
\MIII = \rr_{t}\times\open{-\infty,r_{-}}_r\times\mathbb{S}^{2}_{\theta, \varphi},\
\end{array}
\] 
where $\theta\in \open{0, \pi}$, $\varphi\in \rr/2\pi \zz$ are the spherical coordinates on $\mathbb{S}^{2}$, and the metric in the global {\em Boyer-Lindquist coordinates} $(t, r, \theta, \varphi)$ on  $\MI, \MII$, and $\MIII$ is given by
\[
g=-\left(1-\frac{2Mr}{\rho^2}\right)dt^2-\frac{4aMr\sin^2\theta}{\rho^2}\diff td\varphi +\frac{\rho^2}{\Delta}\diff r^2+\rho^2d\theta^2+\frac{\sigma^2}{\rho^2}\sin^2\theta \diff\varphi^2.
\]
 The time-orientation on $(\MI,g)$ will be fixed by declaring the vector field $-\nabla t=-g^{-1}(dt,\cdot)=\frac{\sigma^2}{\rho^2\Delta}\left(\partial_t+\frac{2Mar}{\sigma^2}\partial_\varphi\right)$, which is timelike on $(\MI,g)$, to be future directed. The time-orientation of ${\rm M}_{\rm II}$ will be inherited from its embedding into ${\rm K}\kst$, see \ref{eclap} below. By \cite[Proposition 5.1]{GHW}, $(\MI, g)$ is globally hyperbolic. 

\subsection{The ${\rm K}\kst$ and $\stk{\rm K}$ spacetimes }\label{turluto} 
To glue the blocks  ${\rm M}_{\rm I}$ and  ${\rm M}_{\rm II}$ together along parts of the black-hole horizon $\{r= r_{+}\}$, one can introduce the {\em Kerr-star} and {\em star-Kerr} coordinates. This will lead to the larger manifolds ${\rm K}\kst$ and $\stk{\rm K}$ introduced below. As the Boyer-Lindquist block $\mathrm{M}_{\rm III}$ does not play a role in our analysis, its corresponding parts will be removed from ${\rm K}\kst$ and $\stk{\rm K}$.

\subsubsection{The ${\rm K}\kst$ spacetime} 
The ${\rm K}\kst$ spacetime is defined as the manifold
\[
{\rm K}\kst= \rr_{t\kst}\times\open{r_{-}, +\infty}_{r}\times \mathbb{S}^{2}_{\theta, \varphi\kst}, 
\]
equipped with the metric
\[
g= g_{tt}dt^{*2}+ 2 g_{t\varphi}dt\kst d\varphi\kst+ g_{\varphi\varphi}d\varphi^{*2}+ 2 dt\kst dr- 2a\sin^{2}\theta d\varphi\kst dr+ \rho^{2}d\theta^{2}
\]
in terms of the global {\em Kerr-star coordinates} $(t\kst, r, \theta, \varphi\kst)$. The coordinate vector fields $\p_{r}$, $\p_{\theta}$ of the Kerr-star coordinates will be denoted by $\p_{r\kst}$, $\p_{\theta\kst}$. The time-orientation of ${\rm K}\kst$ is fixed by declaring the null vector $-\p_{r\kst}$ to be future directed.

\subsubsection{Embedding ${\rm M}_{\rm I}$ and ${\rm M}_{\rm II}$ into ${\rm K}\kst$}\label{eclap}

Define $x(r)$ and $\Lambda(r)$ for $r\in \open{r_{-}, r_{+}}\cup \open{r_{+}, +\infty}$ up to constants by
\beq\label{eq:TLambda}
\frac{dx}{dr}=\frac{r^2+a^2}{\Delta}, \ \ \frac{d\Lambda}{d r}=\frac{a}{\Delta}.
\eeq
 
Then the map  $j\kst: \MI\cup \MII\to {\rm K}\kst$ defined by
\[
t\kst\circ j\kst= t+ x(r), \ \  r\circ j\kst= r, \ \ \theta\circ j\kst= \theta, \ \ \varphi\kst\circ j\kst= \varphi+ \Lambda(r),
\]
can be used to isometrically identify $(\MI, g)$ resp.~ $(\MII, g)$ with $(\MI\kst, g)$, resp.~$(\MII\kst, g)$.  Here,
\[
\MI\kst= {\rm K}\kst\cap\{r_{+}<r\}, \  \ \MII\kst= {\rm K}\kst\cap \{r_{-}<r<r_{+}\},
\]
and the embedding of ${\rm M}_{\rm I}$ and ${\rm M}_{\rm II}$  into ${\rm K}\kst$ respects the time orientation.

Moreover, $j\kst$ allows to glue ${\rm M}_{\rm I}$ with ${\rm M}_{\rm II}$ inside ${\rm K}\kst$ along the {\em future horizon}
\[
\sH_+ = \rr_{t\kst}\times\{ r= r_{+} \} \times \mathbb{S}^{2}_{\theta, \varphi}.
\]

We follow the notation in \cite{GHW}: 

\begin{definition}{\cite[Definition 5.2]{GHW}}\label{defdeMIunionMII}
We set   \[
{\rm M}_{{\rm I}\cup {\rm II}}\defeq (j\kst)^{-1}({\rm K}\kst)= \MI\cup\MII\cup \sH_{+}, 
\]
with the spacetime structure inherited from ${\rm K}\kst$.
\end{definition}
\noindent
 By \cite[Proposition 5.3]{GHW}, the spacetime $({\rm M}_{{\rm I}\cup {\rm II}}, g)$ is globally hyperbolic. 

\subsubsection{The $\stk{\rm K}$ spacetime}
Similarly, the $\stk{\rm K}$ spacetime is defined as the manifold
\[
\stk{\rm K}=  \rr_{\stk t}\times \open{r_{-}, +\infty}_{r}\times \mathbb{S}^{2}_{\theta, \stk\varphi},
\]
equipped with the metric
\[
g= g_{tt}d\stk t^{2}+ 2 g_{t\varphi}d\stk t d\stk\varphi+ g_{\varphi\varphi}d\stk\varphi^{2}- 2 d\stk t dr+ 2a\sin^{2}\theta d\stk\varphi dr+ \rho^{2}d\theta^{2}
\]
in terms of the global {\em star-Kerr coordinates} $(\stk t, r, \theta, \stk\varphi)$. The coordinate vector fields $\p_{r}$, $\p_{\theta}$ of the star-Kerr coordinates are denoted as $\p_{{}^{*}\!r}$, $\p_{{}^{*}\!\theta}$, as before. The time-orientation of $\stk{\rm K}$ is fixed by declaring the null vector $\p_{\stk r}$ to be future directed.

\subsubsection{Embedding ${\rm M}_{\rm I}$ and ${\rm M}_{\rm II}$ into $\stk{\rm K}$}
Again, one can isometrically identify  $(\MI, g)$ resp.~$(\MII, g)$  with   $(\stk\MI, g)$, resp.~$(\stk\MII, g)$ using the map  $\stk j: \MI\cup \MII\to \kst{\rm K}$ defined by
\[
\stk t\circ \stk j= t- x(r), \ \  r\circ \stk j= r, \ \ \theta\circ \stk j= \theta, \ \ \stk \varphi\circ \stk j= \varphi- \Lambda(r).
\]
Here,
\[
\stk\MI= \stk{\rm K}\cap \{r_{+}<r\}, \ \ \stk\MII= \stk{\rm K}\cap \{r_{-}<r<r_{+}\},
\]
and the embedding of ${\rm M}_{\rm I}$ (resp.~${\rm M}_{\rm II}$)  into $\stk{\rm K}$ respects (resp.~reverses) the time orientation.
Furthermore, $\stk j$ allows to glue  ${\rm M}_{\rm I}$ with ${\rm M}_{\rm II}$ inside $\stk{\rm K}$ along the \emph{past horizon}
\beq\label{corr.e2}
\sH_- =\rr_{\stk t}\times\{ r= r_{+} \} \times \mathbb{S}^{2}_{\theta, \stk\varphi}.
\eeq
Note that in the original Boyer--Lindquist coordinates, the future/past horizon $\sH_\pm$ is reached at positive/negative infinite values of $t$.

\subsection{Conformal extension of ${\rm M}_{\rm I}$}\label{confconf}
\begin{figure}[H]
\includegraphics[scale=1]{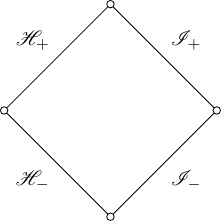}
\caption{The conformal extensions of $\MI$.}\label{fig10}
\end{figure}

Later, we will need the Penrose conformal extension of ${\rm M}_{\rm I}$.
Set $\hat g=w^{2} g$, where $w=r^{-1}\in \open{0, r_{+}^{-1}}$. Then $\hat{g}$ extends as a smooth, non-degenerate metric to $w\in \open{-\infty, r_{+}^{-1}}$, and one can find $\epsilon_{0}>0$ so that $\hat{g}$ is Lorentzian for $w\in \open{-\epsilon_{0}, r_{+}^{-1}}$.
Hence, the conformal extension of ${\rm M}_{\rm I}$ can be defined as
\[
\hat{{\rm M}}_{\rm I}\defeq\rr_{t\kst}\times \opencl{-\epsilon_{0},r_{+}^{-1}}_{w}\times \ss^2_{\theta,\varphi\kst}.
\]
It includes \emph{past null infinity}, i.e. the null hypersurface
\beq\label{corr.e3}
\sI_-=\rr_{t\kst}\times \{ w=0 \} \times \ss^2_{\theta,\varphi\kst}.
\eeq
If, instead, one extends $\hat{g}$ in terms of the coordinates $(\stk t, w, \theta,\stk \varphi)$ to
\[
\rr_{\kst t}\times\open{-\epsilon_{0}, r_{+}^{-1}}_{w}\times \bS^{2}_{\theta \stk \varphi},
\]
one can define \emph{future null infinity} $\sI_+=\rr_{\kst t}\times \{w=0\}\times \bS^{2}_{\theta\stk\varphi}$. In both cases, the time orientation of the conformal extension of ${\rm M}_{\rm I}$ is  inherited from that of ${\rm M}_{\rm I}$.
 Figure \ref{fig10} (compare \cite[Figure 2]{GHW}) shows the resulting conformal diagram.

\subsection{The Kerr-Kruskal extension}\label{subsec.kk}
Finally, we require an extension of the manifold that includes all of the black hole horizon $\{r=r_+\}$. 
To this end, consider the {\em Kerr-Kruskal extension} $({\rm M},g)$ given by the manifold 
\[
{\rm M}= \rr_{U}\times \rr_{V}\times \bS^{2}_{\theta,\varphi^{\t}}.
\]
equipped with the metric  \cite[Proposition 3.5.3]{N}
 \beq\label{trif.5}
\bea
g&=\frac{G^{2}(r)a^{2}\sin^{2}\theta}{4\kappa_{+}^{2}\rho^{2}}\frac{(r-r_{-})(r+r_{+})}{(r^{2}+ a^{2})(r_{+}^{2}+a^{2})}\Big(\frac{\rho^{2}}{r^{2}+a^{2}}+\frac{\rho_{+}^{2}}{r_{+}^{2}+a^{2}}\Big)(U^{2}dV^{2}+ V^{2}dU^{2})\\
&\phantom{=}\,+ \frac{G(r)(r-r_{-})}{2\kappa_{+}^{2}\rho^{2}}\Big(\frac{\rho^{4}}{(r^{2}+a^{2})^{2}}+\frac{\rho_{+}^{4}}{(r_{+}^{2}+a^{2})^{2}}\Big)dUdV\\
&\phantom{=}\,+\frac{G^2(r)a^2\sin^2\theta}{4\kappa_+^2\rho^2}\frac{(r+r_+)^2}{(r_+^2+a^2)^2}\left(UdV-VdU\right)^2\\
&\phantom{=}\,+\frac{G(r)a\sin^{2}\theta}{\kappa_{+}^{2}\rho^{2}(r_{+}^{2}+a^{2})}\big(\rho_{+}^{2}(r-r_{-})+ (r^{2}+a^{2})(r+r_{+})\big)(UdV- VdU)d\varphi^{\sharp}\\
&\phantom{=}\,+ \rho^{2}d\theta^{2}+ \Big(r^{2}+ a^{2}+ \frac{2Mra^{2}\sin^{2}\theta}{\rho^{2}}\Big)\sin^{2}\theta d\varphi^{\sharp 2}.
\eea
 \eeq
 Here, 
\beq
\kappa_{\pm}= \frac{r_{\pm}- r_{\mp}}{2(r_{\pm}^{2}+a^{2})}
\eeq
are the surface gravities of the black hole and Cauchy horizon, $r$ is defined as a smooth function on $M$ via the equation
 \[
\frac{ r- r_{+}}{UV}= G(r), \hbox{ for }G(r)= \e^{- 2\kappa_{+}r}(r-r_{-})^{r_{-}/r_{+}},
 \]
and $\rho_{+}= \rho(r_{+}, \theta)$.
The global coordinates $(U, V, \theta, \varphi^{\t})$ on ${\rm M}$ are called {\em KBL coordinates}.
By \cite[Proposition 5.4]{GHW}, the Lorentzian manifold $({\rm M}, g)$ is globally hyperbolic. 

To isometrically embed ${\rm M}_{\rm I}$, ${\rm M}_{\rm II}$, ${\rm M}_{\rm I}'$, and $ {\rm M}_{\rm II}'$ into ${\rm M}$, respecting the time orientations, one employs the following coordinate changes \cite{GHW}:
 \beq\label{e2}
 \begin{array}{l}
 U= \e^{-\kappa_{+}\stk t}, \ V= \e^{\kappa_{+}t\kst}, \ \hbox{ on }\MI, \\[2mm]
 U= -\e^{-\kappa_{+}\stk t}, \ V= \e^{\kappa_{+}t\kst}\hbox{ on }\MII,\\[2mm]
  U= -\e^{-\kappa_{+}\stk t}, \ V= -\e^{\kappa_{+}t\kst}\hbox{ on }\MI',\\[2mm]
 U= \e^{-\kappa_{+}\stk t}, \ V= -\e^{\kappa_{+}t\kst}\hbox{ on }\MII',\\[2mm]
 \varphi^{\t}= \12(\varphi\kst+ \stk\varphi- \frac{a}{r_{+}^{2}+a^{2}}(t\kst+ \stk t))= \varphi- \frac{a}{r_{+}^{2}+a^{2}}t.
\end{array}
  \eeq
 
In this way, ${\rm M}_{\rm I}$, ${\rm M}_{\rm II}$, ${\rm M}_{\rm I}'$, and $ {\rm M}_{\rm II}'$ are identified with
  \[
  \bea
 &{\rm M}_{\rm I}= \{U>0, \ V>0\}, \ \ {\rm M}_{\rm II}= \{U<0, \ V>0\}, \\ &{\rm M}_{\rm I'}=\{U<0, \ V<0\}, \ \ {\rm M}_{\rm II'}= \{U>0, \ V<0\},
 \eea
 \]
respectively, where we are using a slight alteration of the prime-notation for time-reversed manifolds. The complement in $\MK$ of these four quadrants are the {\em long horizons} $\sH_{R}\defeq\{U=0\}$ and $\sH_{L}\defeq\{V=0\}$. Their intersection is the {\em crossing sphere} $S(r_{+})\defeq \{U= V= 0\}$. 

 \subsection{Killing vector fields}
An important role is played by the two Killing vector fields
\[
\begin{array}{l}
v_{\sI}= \p_{t}= \kappa_{+}(- U \p_{U}+ V\p_{V})-\Omega_{\sH}\p_{\varphi^{\t}}, \\[2.5mm]
  v_{\sH}= \p_{t}+ \Omega_{\sH}\p_{\varphi}= \kappa_{+}(- U \p_{U}+ V\p_{V}),
\end{array}
\]
where $\Omega_{\sH}= \frac{a}{r_{+}^{2}+a^{2}}$  is the \emph{angular velocity} of the horizon. The vector field  $v_{\sH}$ is tangent to  $\sH_{-}$ (in  $\stk{\rm K}$), while $v_{\sI}$ is tangent to $\sI_{-}$ (in the conformal extension $\hat{{\rm M}}_{\rm I}$ of ${\rm M}_{\rm I}$). None of these Killing fields is globally timelike in $\MI$. The \emph{ergoregion} is defined as the regions where $v_{\sI}$ becomes spacelike. We find  
\begin{align}
g(v_{\sI},v_{\sI})>0\quad  \Leftrightarrow \quad \rho^2-2Mr<0.\, 
\end{align}
The Killing field $v_{\sH}$ becomes light-like at the \emph{speed-of-light surface}, and spacelike beyond it, in the region specified by
\begin{align}
\label{eq:ergo 2}
g(v_{\sH},v_{\sH})>0\Leftrightarrow \left(1-\frac{2Mr}{\rho^2}\right)+\frac{4 aMr\sin^2\theta}{\rho^2}\Omega_{\sH}-\frac{\sigma^2}{\rho^2}\sin^2\theta\Omega_{\sH}^2<0.
\end{align}

\subsection{Hamiltonian flow}
In this subsection, we will study the bicharacteristics in $\MI$.  
Let $p(x,\xi)=g_x^{-1}(\xi,\xi)$ and $G=G_r+G_\theta=\rho^{-2}g^{-1}_x(\xi,\xi)$, where \cite{Dy}
\begin{align}
\label{e Gr}
G_r=\Delta\xi_r^2-\frac{1}{\Delta}\left((r^2+a^2)\xi_t+a\xi_\varphi\right)^2,\\
G_\theta=\xi_\theta^2+\frac{1}{\sin^2\theta}\left(a\sin^2\theta\xi_t+\xi_\varphi\right)^2.
\label{e Gtheta}
\end{align}
Bicharacteristics are flow lines of the Hamiltonian vector field $H_p$ contained in $\{p=0\}\setminus o$. 
\begin{definition}
Let $\gamma(s)$ be a maximally extended bicharacteristic. We say that $\gamma(s)$ is trapped as $s\rightarrow +\infty$ if there exist $r_+<r_0<R_0<\infty$ such that $r_0\le r(\gamma(s))\le R_0$ for all $s\ge 0$ (and as a consequence, $\gamma(s)$ exists for all $s\ge 0$). Let $\Gamma^-$ be the union of all trapped geodesics as $s\rightarrow+\infty$. $\Gamma^+$ is defined similarly as all trapped geodesics when $s\rightarrow-\infty$. The trapped set $K$ is defined as $K=\Gamma^+\cap \Gamma^-$.   
\end{definition}
Let $\tilde \xi=(\xi_t,\xi_r,\xi_\varphi,\xi_\theta)$.
By \cite[Proposition 3.2]{Dy}, the trapped set is characterized by
\begin{align}
 K=\{G=\partial_rG_r=\xi_r=0,\,  \tilde\xi\neq 0\}\subset T^*{\rm M}\backslash o.
\end{align}
By \cite[Proposition 3.5]{Dy}, \cite [Proposition 12.2.26]{Mi2} we know that 
\begin{align}
\Gamma^{\pm}=\left\{(r,\hat x;\xi_r,\hat\xi):(\hat x,\hat\xi)\in \hat K; \xi_r=\pm\sgn(r-r^\prime_{\hat x, \hat\xi})\sqrt{\Phi_{\hat x,\hat\xi}(r)\Delta_r^{-1}}\right\}\, ,
\end{align}
where $\Phi_{\hat x,\hat\xi}(r)=-G(r,\hat{x}, 0,\hat{\xi})$, $\hat{K}$ is the projection of the trapped set to $(\hat x; \hat \xi)= (t,\theta,\varphi;\xi_t,\xi_\theta,\xi_\varphi)$ and $r^\prime_{\hat x, \hat\xi}$ is the unique solution of $\Phi_{\hat x,\hat\xi}(r)=0$. It is a submanifold of $T^*(\mathbb{R}\times\mathbb{S}^2)$ of codimension one. 

Let $\pi:T^*\MI\to \MI$ be the base projection.
\begin{lemma}
\label{lemma3.3}
If $(x,\xi)\in(T^*\MI\setminus\zero)\setminus ((\Gamma^-\cap \{\xi_t<0\})\cup (\Gamma^+\cap \{\xi_t>0\}))$, then $\pi(\gamma(s))$ meets $\sH_-$ or $\sI_-$ at some point $y_0$.  
\end{lemma}
\proof
The lemma is an immediate consequence of \cite[Proposition 5.7, Proposition 5.38]{Mi} for $\sH_-$ and $\sI_-$ replaced by $\sH_+$ and $\sI_+$ and $\Gamma^{\pm}$ by $\Gamma^{\mp}$. At $\sH_-$ and $\sI_-$, however, the coordinates used in \cite{Mi} are singular. We therefore use the symmetry $t\leftrightarrow -t,\, \varphi\leftrightarrow -\varphi$ of the metric. This symmetry sends $t\kst,\, \varphi\kst$ to $-\stk t,\, -\stk\varphi$ and therefore $\sH_+$ to $\sH_-$ and $\sI_+$ to $\sI_-$, as well as $\xi_t$ to $-\xi_t$. 
\qed

\section{Traces}
\label{secTraces}
\subsection{Decomposition of $\Sol_{{\rm L}^{2}}(\MI)$}
Following \cite{GHW}, there exists an orthogonal decomposition 
\beq\label{e4.10ccc}
\Sol_{{\rm L}^{2}}(\MI)= \Sol_{{\rm L}^{2}, \sH_{-}}(\MI)\oplus \Sol_{{\rm L}^{2}, \sI_{-}}(\MI),
\eeq
where $\Sol_{{\rm L}^{2}, \sH_{-}}(\MI)={\rm P}_{\sH_{-}}\Sol_{{\rm L}^{2}}(\MI)$, 
$\Sol_{{\rm L}^{2}, \sI_{-}}(\MI)={\rm P}_{\sI_{-}}\Sol_{{\rm L}^{2}}(\MI)$ 
and ${\rm P}_{\sH_{-}/\sI_{-}}$ are suitable projections separating solutions going to the past horizon from those going to past null infinity. These projections are constructed by means of asymptotic velocities, we refer to \cite{GHW}, \cite{HN} for details.

\subsection{Traces on $\sH_{-}$}\label{sec11.7}
Note that in terms of the KBL coordinates defined in Subsect.~\ref{subsec.kk}, $\sH_-$ is given by  $\sH_{-}= \{0\}_{V}\times\open{0, +\infty}_{U}\times \SS^{2}_{\theta, \varphi^{\t}}$.
Let ${\rm L}^{2}(\sH_{-})$ be the completion of $\coinf(\sH_{-}; \cc^{2})$ for the (degenerate) scalar product
\beq\label{e12.7}
(\phi|\phi)_{\sH_{-}}= -\i\int_{\sH_{-}}\bar{\phi}\dual\Gamma(\nabla V)\phi |g|^{\12} \diff U d\theta d\varphi^{\t}.
\eeq
Then the trace on $\sH_-$, which is defined as
\beq\label{corr.e7}
{\rm T}_{\sH_{-}}\phi= \phi_{| \sH_{-}}\in \cinf(\sH_{-};\cc^2)
\eeq
 for $\phi\in \Sol_{\rm sc}(\MI)$, uniquely extends to a bounded operator ${\rm T}_{\sH_{-}}: \Sol_{{\rm L}^{2}}(\MI)\to {\rm L}^{2}(\sH_{-})$ with  
$\Ker {\rm T}_{\sH_{-}}= \Sol_{{\rm L}^{2}, \sI_{-}}(\MI)$ by Gérard et al. \cite[Proposition 7.1]{GHW}. The proof of this proposition also shows that $(\cdot| \cdot)_{\sH_{-}}$ is positive semidefinite. 
Let us also set
\beq\label{e12.1}
\cS_{\sH_{-}}: \Sol_{\rm sc}(\MI)\ni \phi\mapsto \phi\dual \mathfrak{i}_{| \sH_{-}}\in \cinf(\sH_{-}; \cc),
\eeq
so that
\[
\cS_{\sH_{-}}\phi= {\rm T}_{\sH_{-}}\phi\dual \mathfrak{i}.
\]
Here, $\mathfrak{i}$ is one basis vector of a suitable spin frame, we refer to \cite{GHW} for details. 

Let 
\beq\label{e4.10d}
C_{1}= \e^{-\kappa_{+}r_{+}/2}(r_{+}- r_{-})^{M/2r_{+}}, \ \ p_{+}(\theta)= r_{+}+ \i a \cos \theta.
\eeq
As in \cite[Definition 7.2]{GHW}, let $L^{2}(\sH_{-}; \cc)$ denote the closure of $\coinf(\sH_{-}; \cc)$ for the scalar product
  \[
  (\mathfrak{f}_{1}| \mathfrak{f}_{1})= \kappa_{+}^{-1}\dfrac{1}{\sqrt{2}}C_{1}^{2}\int_{\rr^{+}_{U}\times \SS^{2}_{\theta, \varphi^{\t}}}|p_{+}|(\theta)|\mathfrak{f}_{1}|^{2}\sin \theta \diff U d\theta d\varphi^{\t}.
  \]
Then the map $\cS_{\sH_{-}}: \Sol_{{\rm L}^{2}, \sH_{-}}(\MI)\to L^{2}(\sH_{-};\cc )$ is unitary, see \cite[Proposition 7.3]{GHW}.

\subsection{Traces on $\sI_{-}$}\label{sec11.8}
\subsubsection{Conformal rescaling}\label{sec11.8.1}
Following \cite[Subsect.~8.3]{HN}, consider the conformal extension $(\hat{\MI},\hat{g})$ introduced in Subsect.~\ref{confconf}. Objects canonically attached to $(\MI,\hat{g})$ will be decorated with hats. In particular, we have
\begin{equation}
\label{e4.2e}
\hat{\Gamma}(v)= \Gamma(v),\, \hat{\DD}= w^{-3}\DD w.
\end{equation}

\subsubsection{Traces on $\sI_{-}$}
For $\phi\in \Sol_{\rm sc}(\MI)$, the trace on $\sI_-$ is defined by
\beq\label{corr.e8}
{\rm T}_{\sI_{-}}\phi\defeq \hat{\phi}_{| \sI_{-}}, \ \ \hat{\phi}= w^{-1}\phi\in \Sol_{\rm sc}(\hat{\DD}).
\eeq
By \cite[Proposition 7.4]{GHW}, it uniquely extends to a bounded operator ${\rm T}_{\sI_{-}}: \Sol_{{\rm L}^{2}}(\MI)\to {\rm L}^{2}(\sI_{-})$, where ${\rm L}^{2}(\sI_{-})$ is the completion of $\coinf(\sI_{-}; \cc^{2})$ for the scalar product
\[
(\hat{\phi}| \hat{\phi})_{\sI_{-}}=  -\i\int_{\sI_{-}}\bar{\hat{\phi}}\dual\hat{\Gamma}(\hat{\nabla} w)\hat{\phi}|\hat{g}|^{\12} \diff t\stk d\theta d\varphi\stk.
\]
Moreover, one has $\Ker {\rm T}_{\sI_{-}}= \Sol_{{\rm L}^{2}, \sH_{-}}(\MI)$.
The proof of the proposition also shows that $(\cdot| \cdot)_{\sI_{-}}$ is positive semidefinite.
Let us set
\beq\label{e12.2}
\cS_{\sI_{-}}: \Sol_{\sc}(\MI)\ni \phi\mapsto \hat{f}_{0}= \hat{\phi}\dual \hat{\mo}_{| \sI_{-}}\in \cinf(\sI_{-}; \cc),
\eeq
so that
\[
\cS_{\sI_{-}}\phi= {\rm T}_{\sI_{-}}\phi\dual \hat{\mo}.
\]
Here, $\hat{\mo}=\mo$ is one basis vector of a suitable spin frame, we refer to \cite{GHW} for details. 

If we denote by $L^{2}(\sI_{-}; \cc)$ the closure of $\coinf(\sI_{-}; \cc)$ for the scalar product
 \[
 (\hat{f}_{0}| \hat{f}_{0})=\dfrac{1}{\sqrt{2}}\int_{\rr_{t\stk}\times \SS^{2}_{\theta, \varphi\stk}}|\hat{f}_{0}|^{2} \sin \theta \diff t\stk d \theta d \varphi \kst.
 \]
as in \cite[Definition 7.5]{GHW}, then by Gérard et al. \cite[Proposition 7.6]{GHW}, the map $ \cS_{\sI_{-}}: \Sol_{{\rm L}^{2}, \sI_{-}}(\MI)\to L^{2}(\sI_{-}, \sin \theta \diff t\kst d\theta d \varphi\kst)$ is unitary.

We recall 
\begin{proposition}{\cite[Proposition 7.7]{GHW}}\label{thm12.1}
\begin{enumerate}
 \item The map $\cS_{\MI}= \cS_{\sH_{-}}\oplus \cS_{\sI_{-}}$ from $\2Sol(\MI)$
 to $L^{2}(\sH_{-}; \cc)\oplus L^{2}(\sI_{-}; \cc)$ is unitary. 
 \item One has
 \[
 \begin{array}{l}
\cS_{\sH_{-}}\circ \i^{-1}\cL_{\sH}= -\i^{-1}\kappa_{+}(U\p_{U}+ \12)\circ \cS_{\sH_{-}}, \\[2mm]
 \cS_{\sI_{-}}\circ \i^{-1}\cL_{\sI}= \i^{-1}\p_{t\kst}\circ \cS_{\sI_{-}},
\end{array}
\]
\end{enumerate}

 \end{proposition}
 \begin{remark}
 Proposition \ref{thm12.1} is an asymptotic completeness result, we refer to \cite{GHW} and \cite{HN} for details. 
\end{remark}

 \subsection{Decomposition of $\2Sol(\MK)$}
The hypersurface $\Sigma_{\MK}= \{U= V\}$ is a space-like Cauchy surface in $\MK$ by \cite[Proposition C.12]{GHW}, which can be split up into \cite[(C.18)]{GHW}
\[
\Sigma_{\rm I}\defeq \Sigma_{\MK}\cap \MI= \{t=0\}\cap \MI, \ \ \Sigma_{\rm I'}\defeq \Sigma_{\MK}\cap \MIp= \{t=0\}\cap \MIp,
\]
and $S(r_{+})$, the latter being of measure zero in $\Sigma_{\MK}$ for the induced Riemannian metric. Therefore, if $L^{2}(\Sigma)$ denotes the completion of $\coinf(\Sigma; \SS^{*}_{\Sigma})$ for the scalar product $\nu_{\Sigma}$ defined in  \eqref{scalarproduct}, the map
\[
L^{2}(\Sigma_{\MK})\ni f\mapsto f_{| \Sigma_{\rm I}}\oplus f_{| \Sigma_{\rm I'}} \in L^{2}(\Sigma_{\rm I})\oplus L^{2}(\Sigma_{\rm I'}) 
\]
is unitary. By the identification of solutions with their Cauchy data, this yields a unitary map
\beq\label{frip.7}
\2Sol(\MK)\ni \phi\mapsto \phi_{\rm I}\oplus \phi_{\rm I'}\in \2Sol(\MI)\oplus \2Sol(\MIp),
\eeq
where $\phi_{\mathrm{I}}$, resp.~$\phi_{\mathrm{I'}}$ are the restrictions of $\phi$ to $\MI$, resp.~$\MIp$. 

Finally, the identification of $\MI'$ and $\MIp$ as spacetimes through the orientation and time-orientation preserving isometry
\[
R: \MIp\ni (U, V, \theta, \varphi^{\t})\mapsto (-U, -V, \theta, \varphi^{\t})\in \MI'
\]
induces a unitary map  $R: \2Sol(\MIp)\to \2Sol(\MI')$, $$R\phi= \phi \circ R.$$ 

\subsection{Traces on the long horizon and at infinity}
\subsubsection{Traces on the long horizon}
In the sequel, we discuss the long horizon  $\sH_{L}=\{V=0\}$ in $\MK$, which will simply be denoted as $\sH$ for convenience. Then, denoting by ${\rm L}^{2}(\sH)$ the completion of $\coinf( \sH; \cc^{2})$ for the scalar product
\[
(\phi| \phi)_{\sH}= - \i \int_{\sH}\bar{\phi}\dual \Gamma(\nabla V) \phi | g|^{\12}dU d\theta d\varphi^{\t},
\]
one clearly has
\[
{\rm L}^{2}(\sH)\sim {\rm L}^{2}(\sH_{-})\oplus {\rm L}^{2}(\sH_{-}'),
\]
thanks to the decomposition
\[
\sH_{L}= \sH_{-}\cup \sH_{-}'\cup S(r_{+}), \  \ \sH_{-}= \{V=0, \ U>0\}, \ \ \sH_{-}'=\{V= 0, \ U<0\}= R(\sH_{-}).
\]

Therefore, the traces 
\[
{\rm T}_{\sH_{-}}\phi\defeq {\rm T}_{\sH_{-}}\phi_{\rm I}, \ \ {\rm T}_{\sH_{-}'}\phi\defeq R {\rm T}_{\sH_{-}}R \phi_{\rm I'}, 
\]
combine to a trace on the long horizon,
\[
{\rm T}_{\sH}\defeq  {\rm T}_{\sH_{-}}\oplus {\rm T}_{\sH_{-}'}: \2Sol(\MK)\to {\rm L}^{2}(\sH).
\]

Similarly, if we denote by $L^{2}(\sH; \cc)$ the closure of $\coinf(\sH; \cc)$ for the scalar product
 \[
  (\mathfrak{f}_{1}| \mathfrak{f}_{1})= \kappa_{+}^{-1}\dfrac{1}{\sqrt{2}}C_{1}^{2}\int_{\rr_{U}\times \SS^{2}_{\theta, \varphi^{\t}}}|p_{+}|(\theta)|\mathfrak{f}_{1}|^{2}\sin \theta \diff U d\theta d\varphi^{\t},
  \]
the decomposition of the long trace implies
\[
L^{2}(\sH; \cc)\sim L^{2}(\sH_{-}; \cc)\oplus L^{2}(\sH_{-}'; \cc),
\]
and the traces
 \[
 \cS_{\sH_{-}}\phi\defeq \cS_{\sH_{-}}\phi_{\mathrm{I}}, \ \ \cS_{\sH_{-}'}\phi\defeq R \cS_{\sH_{-}} R\phi_{\rm I'}, 
 \]
 combine to a trace on the long horizon,
 \[
 \cS_{\sH}\defeq \cS_{\sH_{-}}\oplus \cS_{\sH_{-}'}: \2Sol(\MK)\to L^{2}(\sH; \cc).
 \]
 
 \subsubsection{Traces at infinity}
 For $\sI_-$ and $\sI_{-}'$, we simply set
 \[
 {\rm T}_{\sI_{-}}\phi\defeq {\rm T}_{\sI_{-}}\phi_{\rm I}, \ \ {\rm T}_{\sI_{-}'}\phi\defeq {\rm T}_{\sI_{-}}R\phi_{\rm I'},
 \]
 \[
\cS_{\sI_{-}}\phi\defeq \cS_{\sI_{-}}\phi_{\rm I}, \ \ \cS_{\sI_{-}'}\phi\defeq \cS_{\sI_{-}}R\phi_{\rm I'}.
 \]

 \subsubsection{Summary}  The following theorem is \cite[Theorem 7.8]{GHW}, it summarizes the construction. 
\begin{theorem}
The map
  \beq\label{frip.-3}
 \begin{array}{l}
 \cS_{\MK}\defeq \cS_{\sH}\oplus \cS_{\sI_{-}}\oplus\cS_{\sI_{-}'}: \2Sol(\MK)\to L^{2}(\sH; \cc)\oplus L^{2}(\sI_{-}; \cc)\oplus L^{2}(\sI_{-}'; \cc)
 \end{array}
 \eeq
is unitary. 
\end{theorem}

\begin{figure}[H]
\includegraphics[scale=1]{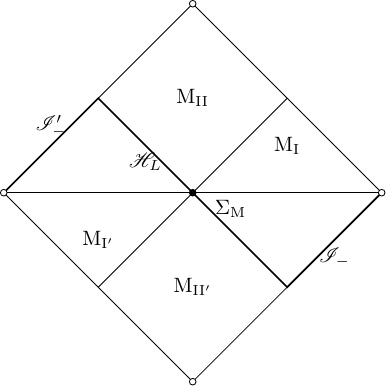} 
 \caption{The Kerr-Kruskal spacetime with  Cauchy surface $\Sigma_{\MK}$.}\label{fig9}
 \end{figure}

The geometric situation is illustrated in Figure \ref{fig9} (compare \cite[Figure 4]{GHW}).
 
 \begin{remark}
 \begin{enumerate}
 \item The constructions of this section are based on the scattering theory developed in \cite{HN}. Whereas the operators $T_{\sH_-}$ and $T_{\sI_-}$ are defined without a reference to a tetrad, for the definition of $S_{\sH_-}$ and $S_{\sI_-}$ we use tetrads which are suitable renormalizations of Kinnersley's tetrad to make them smooth up to the past horizon resp. past null infinity.   
 \item Using a tetrad, the Dirac equation can be written as 
 \begin{align*}
 \partial_t\Psi-iH\Psi=0
 \end{align*}
 with a self-adjoint operator $H$ on a suitable Hilbert space. Asymptotic velocities as constructed in \cite{HN} are then self-adjoint operators $P^{\pm}$ such that 
 \begin{align*}
 J(P^{\pm})=s-\lim_{t\rightarrow\pm \infty}e^{-itH}J\left(\frac{r_*}{t}\right)e^{itH}
 \end{align*}
 for all continuous functions $J$ going to zero at infinity. 
 \end{enumerate}
 \end{remark}
 
\section{The Unruh state in in the Kerr-Kruskal spacetime}
\label{secmain}
\subsection{Hadamard states for the Weyl equation}
States are usually defined by their spacetime covariances. It turns out that for the Weyl equation, it is more useful to define the state $\omega$ by  its {\em solution space covariances},  i.e.~operators $C^{\pm}\in B(\2Sol(M))$ such that
\begin{equation}
\label{frip.1}
C^{\pm}\geq 0, \ \ C^{+}+ C^{-}= \one.
\end{equation}
For this paper, we only need the definition in terms of solution space covariances. We refer to \cite{GHW} for the definition in terms of spacetime covariances and the link between both. It turns out that for the Weyl equation, a characterisation of Hadamard states in terms of wave front sets of $L^2$ solutions can be given, see \cite[Theorem 4.3]{GHW}:
\begin{theorem}\label{propcoup.2}
 Suppose that
 \beq\label{ecoup.6}
 \WF((C^{\pm})^{\12}\phi)\subset \cN^{\pm} \ \ \forall \phi\in \Sol_{{\rm L}^{2}}(M).
 \eeq
 Then the state $\omega$  is a Hadamard state.
\end{theorem}

\begin{definition}\label{def13.1}
 The {\em Unruh state} $\omega_{\MK}$ is the quasi-free state on $\CAR(\MK)$  with solution space covariances:
 \beq\label{sloubi2}
 C^{\pm}_{\MK}=  \cS_{\MK}^{-1}\left( \one_{\rr^{\pm}}(-\i^{-1}\p_{U})\oplus \one_{\rr^{\pm}}(\i^{-1}\p_{t\kst})\oplus \one_{\rr^{\mp}}(\i^{-1}\p_{t\kst})\right)\cS_{\MK}.
 \eeq
\end{definition}
We refer to \cite{GHW} and references therein for the definition of the $\CAR(\MK)$ algebra. We also recall the two main results of \cite{GHW}:
\begin{theorem}\label{thm13.2}
\ben
\item The Unruh state $\omega_{\MK}$ is a pure state.
\item The restriction $\omega_{\MIUII}$ of $\omega_{\MK}$ to $\MIUII$ is a pure state.
\item The restriction $\omega_{\rm M_{I}}$ of $\omega_{\MK}$  to $\MI$ has   covariances
 \[
 C^{\pm}_{\MI}= \cS_{\MI}^{-1}\left(\chi^{\pm}_{\sH_{-}}(-\i^{-1}\kappa_{+}(U\p_{U}+ \12))\oplus \chi^{\pm}_{\sI_{-}}(\i^{-1}\p_{t\kst})\right)\cS_{\MI}
 \]
 for
 \beq\label{ezlib.2}
 \chi^{\pm}_{\sI_{-}}(\lambda)= \one_{\rr^{\pm}}(\lambda), \ \ \chi^{\pm}_{\sH_{-}}(\lambda)= \big(1+ \e^{\mp T_{\rm H}^{-1}\lambda}\big)^{-1},
 \eeq
 where $T_{\rm H}= (2\pi)^{-1}\kappa_{+}$ is the  \emph{Hawking temperature}. 
 \item The restriction $\omega_{\rm M_{I'}}$ of $\omega_{\MK}$ to $\MIp$ is  the image of $\omega_{\rm M_{I}}$ under $R$, and has   covariances
 \[
C_{\MIp}^{\pm}=  R\circ C_{\MI}^{\mp}\circ R. 
 \]
  \een
\end{theorem}
\begin{theorem}\label{thm13.1}
There exists $0<a_{0}\leq 1$ such that  if $|a|M^{-1}<a_{0}$, then 
 the restriction $\omega_{\MIUII}$ of the Unruh state $\omega_{\MK}$ to $\MIUII$ is a Hadamard state.
\end{theorem}
The hypothesis that $a$ is small is crucial in the proof of Theorem \ref{thm13.1}, but is not needed in Theorem \ref{thm13.2}. The main result of this note states that Theorem \ref{thm13.1} holds in the full subextreme range. 
\begin{theorem}\label{thm13.3}
Theorem \ref{thm13.1} holds for all $0<\abs{a}<M$. 
\end{theorem}

\section{Proof of Theorem \ref{thm13.3}}
\label{secproof}
The only place where the smallness assumption on $\abs{a}$ is used in the proof of Theorem \ref{thm13.1} in \cite{GHW} is in the proof of Proposition 8.7., claiming that $\omega_{\MI}$ is a Hadamard state. Thus, $\omega_{\MIUII}$ being a Hadamard state for all $\abs{a}<M$ follows from proof in \cite[Section 8.5]{GHW}, once we have shown

\begin{proposition}\label{propzlib.1}
Let $0<\abs{a}<M$. Then $\omega_{\MI}$ is a Hadamard state.
\end{proposition}
By Theorem \ref{propcoup.2}, it suffices to prove that 
\begin{equation}
\label{ecoup.66}
\WF((C_{\MI}^{\pm})^{\12}\phi)\subset \cN^{\pm}, \ \forall \phi\in \Sol_{{\rm L}^{2}}(\MI).
\end{equation}
 We will prove only the $+$ case, the $-$ case being analogous. Let  $q_{0}=(x_{0}, \xi_{0})\in \WF((C_{\MI}^+)^{\12}\phi)$ and let $\gamma$ be the null bicharacteristic in $T^{*}\MK$ from $q_{0}$. Let $\pi:T^*\MK\to\MK$ the base projection. In \cite{GHW}, three cases have been distinguished:

{\it Case 1:} $\pi(\gamma)$ intersects $\sI_{-}$  at some point $y_{0}$.  

{\it Case 2:} $\pi(\gamma)$ intersects $\sH_{-}$  at some $y_{0}$.  

{\it Case 3:} $\pi(\gamma)$ doesn't intersect $\sH_{-}$ or $\sI_{-}$. 

Only case 3 required the smallness assumption. Therefore, the only thing to do is to treat case 3 without this assumption. By Lemma \ref{lemma3.3}, case 3 corresponds to $q_0\in (\Gamma^-\cap \{\xi_t<0\})\cup (\Gamma^+\cap \{\xi_t>0\})$.  In the following, we will only consider $\MI$. 

\subsection{Preparations}
\label{sec6.1}
The key formula we will use is (see \cite[Equation (8.7)]{GHW}):
\begin{align*}
C^{\pm}_{\MI}= {\rm P}_{\sH_{-}}\circ \chi^{\pm}_{\sH_{-}}(\i^{-1}\cL_{\sH})+ {\rm P}_{\sI_{-}} \circ \chi^{\pm}_{\sI_{-}}(\i^{-1}\cL_{\sI}).
\end{align*}
An important fact is that the projections ${\rm P}_{\sH_{-}}$ and ${\rm P}_{\sI_{-}}$ commute with functions of $\i^{-1}\cL_{\sH}$ and $\i^{-1}\cL_{\sI}$. We obtain 
\begin{align}
\label{keyformula}
(C^{+}_{\MI})^{1/2}\phi= (\chi_{\sH_{-}}^{+})^{\12}(\i^{-1}\cL_{\sH}){\rm P}_{\sH_{-}}\phi+ (\chi_{\sI_{-}}^{+})^{\12}(\i^{-1}\cL_{\sI}){\rm P}_{\sI_{-}}\phi\eqdef \phi_{\sH}+ \phi_{\sI}.
\end{align}
Note that if $\phi \in\Sol_{{\rm L}^{2}}(\MI)$, so are ${\rm P}_{\sH_{-}}\phi$ and ${\rm P}_{\sI_{-}}\phi$. The following lemma is equivalent to \cite[Lemma 8.6]{GHW}.
\begin{lemma}\label{key-lemma2}
 Let  $X=v_{\sH}$ or $v_{\sI}$ and recall that $\i^{-1}\cL_{X}$ is the self-adjoint generator of a unitary group on $\Sol_{{\rm L}^{2}}(\MI)$. Let  $\chi^{\pm}\in L^\infty(\rr)$ be such that $\chi^\pm- \one_{\rr^{\pm}}\in O(\langle \lambda\rangle^{-\infty})$ and $\sing\supp \chi^\pm\subset \rr^\pm\cup\{0\}$ is compact. Then one has:
\[
\WF(\chi^{\pm}(\i^{-1}\cL_{X})\phi)\subset \{(x,\xi);\, \pm X\dual\xi>0\}\, \hbox{for all }\, \phi\in \Sol_{{\rm L}^{2}}(\MI).
\]
  \end{lemma}
Note that similar results have been obtained in \cite[Theorem 8.4.8]{Hö}, \cite[Theorem 2.8]{SVW} and \cite{PSV}.	

\proof 
  The proof is essentially the same as for \cite[Lemma 8.6]{GHW}. We repeat it here for the convenience of the reader. 
We consider the case of $X=v_{\sI}=\p_t$, the other case being analogous. Let us denote $\phi^+=\chi^+(\i^{-1}\cL_{X})\phi$. Let $q_0=(x^0,\xi^0)=(t^0,r^0,\varphi^0,\theta^0,\xi_t^0,\xi_r^0,\xi_{\varphi}^0,\xi_{\theta}^0)$ with $\xi^0_t<0$.  

Let $\vartheta\in \cinf(\rr)$ be such that $\vartheta (\xi_t^0)=1$, $\vartheta(\lambda)=1$ for $\lambda \ll 0$ and $\vartheta(\lambda)=0$ for $\lambda> 0$ and  $\lambda \in \sing\supp\chi^+$. Setting  $\chi_\infty= \vartheta \chi^+ \in O(\langle \lambda\rangle^{-\infty})$, by functional calculus we have
\beq\label{eq:chizero}
\vartheta(\i^{-1}\cL_{X})\phi^+ = \chi_{\infty} (\i^{-1}\cL_{X})\phi.
\eeq
In coordinates, $\cL_{X} \phi= M(t)\p_t M(t)^{-1} \phi$ for some smooth family of invertible fibre endomorphisms $M(t)$. It follows that 
\[
\vartheta(\i^{-1}\cL_{X})\phi^+ =  M(t) \vartheta(\i^{-1}\p_t) M(t)^{-1}  \phi^+.
\]
Let $A\in\Psi^0$ be a properly supported pseudo-differential operator (in the sense of the usual calculus on manifolds), elliptic at $q_0$ and micro\-supported in a small conic neighbourhood $\Gamma_{0}$ of $q_0$. Since $\xi_t^0<0$, the symbol of $\vartheta(\i^{-1}\p_t)$ equals $1$ on $\Gamma_{0}$ away from $\zero$. Therefore, the operator
\[
B\defeq A \circ M(t) \vartheta(\i^{-1}\p_t) M(t)^{-1}
\]
belongs to $\Psi^0$ and is elliptic at $q_0$, and similarly one can show that $A \circ \chi_{\infty}(\i^{-1}\p_t)$ belongs to $\Psi^{-\infty}$. Thus, by acting on both sides of \eqref{eq:chizero} with $A$ we find $B \phi^+ \in \cinf$. Consequently, $q_0\notin \wf(\phi^+)$. 

The proof of the minus sign version of the statement is analogous. \qeds

\subsection{Characterisation of $\cN^+$}
\label{sec6.2}

Let us consider a generic null covector $\xi\in T_x^*\MI$ for some $x\in\MI$. Then one has the following general result
\begin{lemma}
\label{lemma6.1}
Let $x\in \MI$ and $\xi\in T^*_x\MI$ a null covector. Then $\xi$ is future pointing iff $\xi\cdot (-\nabla t)> 0$.
\end{lemma}
\proof
 We use abstract index notation. $\xi_a$ is future directed if and only if $-\xi^b=-g^{ab}\xi_a$ is future directed if and only if $\bar{C}_x(-\xi^b)=\bar{C}_x(-\nabla t)$, where $\bar{C}_x(u)$ is the causal cone at $x$ containing $u$. By \cite[Lemma 5.29, Exercise 5.3]{ON}, this holds if and only if $\xi_a(-\nabla t)^a=g_{ab}\xi^a(-\nabla t)^b>0$.  
\qeds

$\xi$ must satisfy  $g^{-1}(\xi,\xi)=0$, and therefore  $G=G_r+G_\theta=0$, where $G_r$ and $G_\theta$ are as given in \eqref{e Gr} and \eqref{e Gtheta}.
It can for example be specified by picking any $(\xi_r,\xi_\theta,\xi_\varphi)\in \rr^3$ arbitrary, and setting
\begin{align}
\label{eq:xit-rel}
\xi_t=-\frac{2Mar}{\sigma^2}\xi_\varphi\pm \sqrt{\frac{\Delta}{\sigma^2}\left(\Delta \xi_r^2+\xi_\theta^2+\frac{\rho^4}{\sigma^2\sin^2\theta}\xi_\varphi^2\right)}\, .
\end{align}
Note that the root is non-vanishing, unless either $\Delta=0$ or $(\xi_r,\xi_\theta,\xi_\varphi)=0$, and hence also $\xi_t=0$.
\begin{lemma}
\label{lemma6.3}
If we choose the $+$ sign in \eqref{eq:xit-rel}, then $\xi$ is future pointing. If we choose the $-$ sign, then $\xi$ is past pointing. 
\end{lemma}
\proof
Let $\xi_t$ be determined as above.
By Lemma \ref{lemma6.1}, it suffices to check that $\xi \cdot (-\nabla t)>0$. Then
\begin{align*}
\xi \cdot (-\nabla t)= \frac{\sigma^2}{\rho^2\Delta}\xi_t+\frac{2Mar}{\rho^2\Delta}\xi_\varphi=\pm\frac{\sigma^2}{\rho^2\Delta} \sqrt{\frac{\Delta}{\sigma^2}\left(\Delta \xi_r^2+\xi_\theta^2+\frac{\rho^4}{\sigma^2\sin^2\theta}\xi_\varphi^2\right)}\, .
\end{align*}
Therefore, $\pm\xi \cdot (-\nabla t)>0$ if one chooses the $\pm$ sign in \eqref{eq:xit-rel}.

Since the coordinates used above do not cover the axis of rotation, we have to consider it separately. Due to the symmetry of the spacetime under reflection along the equatorial plane, it is sufficient to consider the north pole. To do so, we use stereographic coordinates on the sphere, given by
\begin{align*}
x_1=\sin\theta \cos\varphi \quad x_2=\sin\theta \sin\varphi\,,\quad 
\xi_\theta=\cot\theta(x_1\xi_1+x_2\xi_2)\, ,\quad \xi_\varphi=x_1\xi_2-x_2\xi_1\,.
\end{align*}
Following the computation in \cite{Dy}, one finds that on the axis, and in this new coordinate system, one can express the condition $G=0$ as
\begin{align*}
&\Delta \xi_r^2-\frac{(r^2+a^2)^2}{\Delta}\xi_t^2+\xi_1^2+\xi_2^2=0\\
\Leftrightarrow &\xi_t=\pm\sqrt{\frac{\Delta}{(r^2+a^2)^2}\left(\xi_1^2+\xi_2^2+\Delta \xi_r^2\right)}\, .
\end{align*}
Similarly, one finds that on the axis $-\nabla t=\frac{r^2+a^2}{\Delta}\partial_t$, since $\partial_\varphi=x_1\partial_{x_2}-x_2\partial_{x_1}$ vanishes there. Therefore, 
\begin{align*}
\xi \cdot (-\nabla t) = \pm\frac{r^2+a^2}{\Delta}\sqrt{\frac{\Delta}{(r^2+a^2)^2}\left(\xi_1^2+\xi_2^2+\Delta \xi_r^2\right)}\, .
\end{align*}
This is positive (negative) if one chooses the upper (lower) sign.
\qeds

 It is interesting to note that
\begin{align}
\frac{\Omega_{\sH}}{a}-\frac{2Mr}{\sigma^2}&\ge\frac{1}{r^2+a^2}-\frac{2Mr}{\sigma^2}\nonumber\\
&=\frac{1}{\sigma^2(r^2+a^2)}\left(\rho^2(r^2+a^2)+2a^2Mr\sin^2\theta-2Mr(r^2+a^2)\right)\nonumber\\
\label{eq:O0 est}
&=\frac{\rho^2\Delta}{\sigma^2(r^2+a^2)}
\end{align}
which is positive if $\Delta>0$.

\subsection{Analysis on the trapped set}
\label{sec6.3}
Let us now consider the trapped set. Recall that the trapped set is characterized by 
\begin{align*}
 K=\{G=\partial_rG_r=\xi_r=0,\, \tilde\xi\neq 0\}\subset T^*{\rm M}\backslash o.
\end{align*}

Assume that we have chosen $\xi$ with $\xi_r=0$ and $\xi_t$ given by one of the two options in \eqref{eq:xit-rel}, so that $G=0$ is satisfied. Note that $(r^2+a^2)\xi_t+a\xi_{\varphi}=0,\, G=0$ entails $\tilde{\xi}=0$. The condition $\partial_rG_r=0,\, \tilde{\xi}\neq 0$ can then be rewritten in the following way \cite{Dy}:
\begin{align}
\label{6.47}
4r\Delta \xi_t-\partial_r\Delta ((r^2+a^2)\xi_t+a\xi_\varphi)&=0\nonumber\\
\Leftrightarrow (2r\Delta-(r-M)(r^2+a^2))\xi_t-(r-M)a\xi_{\varphi}&=0\nonumber\\
\Leftrightarrow \xi_t\left[r^2(r-3M)+a^2(r+M)\right]- a\xi_\varphi (r-M)&=0.
\end{align}
Note that on the trapped set $\xi_t=0$ entails $\tilde \xi=0$. We therefore suppose $\xi_t\neq 0$ in the following.
Multiplying \eqref{6.47} by $\xi_t$ in the first line, one obtains the condition
\begin{align}
\label{2.16}
\xi_t(\xi_t+\Omega_0\xi_\varphi)>0\, ,
\end{align}
where we have set $\Omega_0(r)=a/(r^2+a^2)$. 
\begin{lemma}
\label{lemma6.5}
$\{\xi_t<0\}\cap\{\xi\cdot v_\sH>0\}\cap K=\emptyset.$ 
\end{lemma}
\proof
Note that on the axis of rotation, in the coordinate system $(t,r,x_1,x_2)$ discussed above, one finds $v_\sH=\partial_t$, and therefore on the axis the condition $\xi\cdot v_\sH>0$ reduces to $\xi_t>0$. We can thus work away from the axis of rotation. 

The only way for $(x,\xi)$ to be in the intersection $\{\xi_t<0\}\cap\{\xi\cdot v_\sH>0\}$ is by having $-\Omega_\sH\xi_\varphi<\xi_t<0$, and therefore  $a\xi_\varphi>0$, since $a^{-1}\Omega_\sH>0$. If we also demand that $(x,\xi)\in K$, by \eqref{eq:O0 est} and \eqref{2.16}, we need to take the minus-sign for the root in \eqref{eq:xit-rel}, with $\xi_\theta$ chosen small enough so that 
\begin{align}
\label{eq:ergo omega}
\Omega(\xi_\theta):=\abs{\frac{\xi_t}{a\xi_\varphi}}=\frac{2Mr}{\sigma^2}+\sqrt{\frac{\Delta}{\sigma^2}\left(\frac{\rho^4}{\sigma^2 a^2\sin^2\theta}+\frac{\xi_\theta^2}{a^2\xi_\varphi^2}\right)}<\frac{\Omega_\sH}{a}\, .
\end{align} 
To exclude this possibility, we need to take into account the trapped set condition $\partial_r G_r=0$ in more detail. Starting from \eqref{6.47} and considering that $\xi_t$ and $a\xi_\varphi$ are of opposite sign, one finds
\begin{align}
\label{eq:drGr}
&\xi_t\left[r^2(r-3M)+a^2(r+M)\right]-a\xi_\varphi  (r-M)=0\\\nonumber
\Leftrightarrow &-\Omega(\xi_\theta)\left[r^2(r-3M)+a^2(r+M)\right]-(r-M)=0\\\nonumber
\Leftrightarrow &-\left[r^2(r-3M)+a^2(r+M)\right]=\frac{(r-M)}{\Omega(\xi_\theta)}.
\end{align}
Since $(r-M)>0$ in $\MI$, and by \eqref{eq:ergo omega}, one obtains the estimate
\begin{align}
\label{eq:cond bad case}
&-\left[r^2(r-3M)+a^2(r+M)\right]>\frac{a(r-M)}{\Omega_\sH}\\\nonumber
\Leftrightarrow &P(r):=r^3-3Mr^2+(r_+^2+2a^2)r-Mr_+^2<0.
\end{align}
Using that $r^2-2Mr>-a^2$ in $\MI$, one obtains
\begin{align*}
\partial_r P(r)=3r^2-6Mr+r_+^2+2a^2>r_+^2-a^2=2(M^2-a^2+M\sqrt{M^2-a^2})>0\, ,
\end{align*}
so $P(r)$ is strictly monotonically increasing on $\MI$ as long as $\vert a\vert<M$. Moreover,
\begin{align*}
P(r_+)=r_+^3-3Mr_+^2+r_+^3+2a^2r_+-Mr_+^2=2r_+\Delta(r_+)=0\,.
\end{align*}
Therefore, $P(r)> 0$ on $\MI$, in contradiction to \eqref{eq:cond bad case}. This concludes the lemma.
\qed

\subsection{The backward/forward trapped sets}
\label{sec6.4}
Recall from Section \ref{secKerr} that the backward/forward trapped sets are described by the condition
\begin{align*}
\Gamma^{\pm}=\left\{(r,\hat x;\xi_r,\hat\xi):(\hat x,\hat\xi)\in \hat K; \xi_r=\pm\sgn(r-r^\prime_{\hat x, \hat\xi})\sqrt{\Phi_{\hat x,\hat\xi}(r)\Delta_r^{-1}}\right\}. 
\end{align*}
Here, $\Phi_{\hat x,\hat\xi}(r)=-G(r,\hat{x}, 0,\hat{\xi})$, $\hat{K}$ is the projection of the trapped set to $(t,\theta,\varphi;\xi_t,\xi_\theta,\xi_\varphi)$ and $r^\prime_{\hat x, \hat\xi}$ is the unique solution of $\Phi_{\hat x,\hat\xi}(r)=0$. $\hat{K}$ is a submanifold of $T^*(\mathbb{R}\times\mathbb{S}^2)$ of codimension one. 
The condition on $\xi_r$ ensures that $(r,\hat x;\xi_r,\hat\xi)$ in the backward/forward trapped set is null, i.e. satisfies a relation of the form \eqref{eq:xit-rel}. 
\begin{proposition}
\label{proposition6.7}
If $\xi\cdot \partial_t>0$ or $\xi\cdot v_\sH>0,\, (x,\xi)\in \Gamma^{\pm}$, then  $\xi$ is future-pointing. 
\end{proposition}
\proof
Assume $(x,\xi)=(r,\hat x,\xi_r,\hat \xi)\in\Gamma^{\pm}$.
Let us start with the case $\xi_t<0$, but $\xi\cdot v_\sH>0$. This amounts to $\xi_t>-\Omega_\sH \xi_\varphi$. Since $(\hat x, \hat \xi)\in \hat K$, there must be a radius $\hat{r}$ such that $(\hat r ,\hat x,0, \hat \xi)$ is in the trapped set. However, by Lemma \ref{lemma6.5}, no such $\hat{r}$ exists. 

It then remains to consider the case $\xi_t>0$. Since $\xi\in T^*_x\MI$ is null, $\xi_t$ is given by \eqref{eq:xit-rel} evaluated at $r$. Assume that $\xi$ is past pointing. Then, by Lemma \ref{lemma6.3}, $\xi_t$ is given by the lower sign in \eqref{eq:xit-rel}. Together with $\xi_t>0$, this implies that $a\xi_\varphi<0$ and leads to the upper bound
\begin{align}
\label{eq:upper bound}
 \xi_t<\frac{2Mr}{\sigma^2}\abs{a\xi_\varphi}<\frac{1}{r^2+a^2}\abs{a\xi_\varphi}<\frac{\Omega_\sH}{a}\abs{a\xi_\varphi}\, ,
\end{align}
by an application of \eqref{eq:O0 est}. 

Next, we focus on the condition for $(\hat x,\hat \xi)\in \hat K$. Let us assume that $\hat r$ is the radius so that $(\hat r,\hat x,0,\hat \xi)\in K$. Then we must have $\partial_{ r}G_{ r}(\hat r)=0$. However, given the bound \eqref{eq:upper bound} and the fact that $\xi_t$ and  $a\xi_\varphi$ are of opposite sign, it follows from the proof of Lemma \ref{lemma6.5} that no such $\hat r $ exists. Therefore, this case is excluded. In other words: any $(x,\xi)\in  \Gamma^{\pm}$ that satisfies $\xi \cdot \partial_t>0$ or $\xi\cdot v_\sH>0$ is future pointing.
\qed

\subsection{Proof of Proposition \ref{propzlib.1}}
\label{sec6.5}
Let $(x,\xi)\in \WF((C^{\pm}_{\MI})^{\12}\phi),\, \phi\in \Sol_{{\rm L}^{2}}({\rm M}_{\rm I})$. From Lemma \ref{lemma3.3} and the remarks at the beginning of this section it is clear that we only have to consider the case $(x,\xi)\in \Gamma^{\pm}$. From the key formula \eqref{keyformula} and Proposition \ref{key-lemma2} we see that either $\xi\cdot \partial_t>0$ or $\xi\cdot v_{\sH}>0$. Thus, $\xi$ is future pointing by Proposition \ref{proposition6.7}. As the other cases are already treated, the Unruh state is Hadamard by Theorem \ref{propcoup.2}.
\qed
\section*{Conflict of interests-Data}
On behalf of all authors, the corresponding author states that there is no conflict of interest. Our manuscript has no associated data.

\end{document}